\documentclass[aps,prl,twocolumn,]{revtex4}  % for review and submission

\usepackage{graphicx}  % needed for figures
\usepackage{dcolumn}   % needed for some tables
\usepackage{bm}        % for math
\usepackage{amssymb}   % for math

\usepackage{amsmath}

\usepackage{epstopdf}
\newcommand{\bee}{\begin{equation}}
\newcommand{\ee}{\end{equation}}
\newcommand{\beea}{\begin{eqnarray}}
\newcommand{\eea}{\end{eqnarray}}

\begin{document}
\title{Infrared fixed point of the 12-fermion SU(3)  gauge model based on 2-lattice MCRG matching}
\author{ Anna Hasenfratz}
\email{anna@eotvos.colorado.edu}

\affiliation{Department of Physics, University of Colorado, Boulder, CO-80309-390}

\begin{abstract}
I investigate an SU(3) gauge model with  12 fundamental fermions.  The physically interesting region of this strongly coupled system can be  influenced by an ultraviolet fixed point due to lattice artifacts. I suggest to use a gauge action with an additional negative adjoint plaquette term that  lessens this problem. I also introduce a new analysis method for the 2-lattice matching Monte Carlo renormalization group technique that   significantly reduces finite volume effects. The combination of these two improvements allows me to measure the bare step scaling function in a  region of the gauge coupling where it is clearly negative, indicating a positive renormalization group $\beta$ function and infrared conformality.
\end{abstract}

\maketitle

%\section{Introduction}

Gauge models with many fermions or fermions in higher representations can develop a conformal phase characterized by the emergence of an infrared fixed point  (IRFP) in the gauge coupling. Both the conformal systems and those that are  still chirally broken but are very near to the conformal window could be relevant for physics beyond the standard model. During the last few years many of these models were studied using various lattice simulation techniques \cite{Appelquist:2007hu,DelDebbio:2008zf,Fodor:2008hm,Appelquist:2009ty, Jin:2009mc,Hasenfratz:2009ea,Deuzeman:2009mh,Hasenfratz:2010fi,Pica:2010zz,DeGrand:2010na,DeGrand:2011qd,Fodor:2011tu}. The theory with SU(3) gauge fields and 12 flavors of fundamental fermions has been the subject of extensive investigations, but its infrared behavior is still controversial. Refs. \cite{Appelquist:2007hu,Appelquist:2009ty} used the Schrodinger functional approach to numerically calculate the renormalization group $\beta$ function and concluded that the theory has an IRFP, i.e. it is conformal. Ref.\cite{Deuzeman:2009mh} considered the system at finite temperature and reached similar conclusions. Both of these works used unimproved or only moderately improved actions at strong gauge couplings where lattice artifacts could seriously effect the results. At the same time studies of spectral quantities appeared to be more consistent with a chirally broken system\cite{ Fodor:2008hm, Jin:2009mc}. Early studies using Monte Carlo renormalization group (MCRG) techniques were not able to push deep enough into the strong coupling and remained inconclusive\cite{ Hasenfratz:2009ea,Hasenfratz:2010fi}. Recently a large scale study  \cite{Fodor:2011tu} concluded that high precision data of spectral quantities prefer the chirally broken interpretation, but other groups interpret the same data as more compatible with the conformal behavior\cite{Appelquist:2011dp}. 

In this work I revisit the 12 flavor SU(3) system using MCRG methods. Due to two improvements, one in the lattice action, the other in analyzing the MCRG data, I am able to cover a wider coupling range and can demonstrate that in the investigated region the renormalization group $\beta$ function (actually its lattice analogue, the bare step scaling function) has the opposite  sign of an asymptotically free theory, signaling the existence of an infrared fixed point and  the conformality of the system. 

The basic observation that led to the modified action is the existence of an ultraviolet fixed point due to strong coupling lattice artifacts.  It is well known that the pure gauge SU($N_c$) theory both with $N_c=2$ and 3   exhibits a first order phase transition in the fundamental-adjoint plaquette gauge
action space \cite{Bhanot:1981eb,Bhanot:1981pj}.  This line ends in a second order point that has a (most likely trivial) ultraviolet fixed point (UVFP). For notational convenience I call this new fixed point UVFP-2, while I use G-FP to refer to the perturbative Gaussian fixed point at zero gauge coupling. While the first order phase transition and the UVFP-2 are lattice artifacts and independent of the G-FP and the continuum limit defined there, their existence can strongly influence, even completely change, the scaling behavior of the lattice model.  

Ref. \cite{Hasenbusch:2004yq} studied the scaling of several observables of the SU(3) fundamental-adjoint pure gauge system with  adjoint coupling $\beta_A=0$, -2.0 and -4.0, far away from the endpoint of the first order line that occurs around $\beta_A\approx 2.0$.  
Nevertheless the data showed very large scaling violations, even lack of scaling, at couplings 
near the extension of the first order phase transition line. 
As expected, the scaling violations decrease with negative adjoint terms in the action, i.e. farther  from the second order endpoint.   
Refs. \cite{Tomboulis:2007re, Tomboulis:2007rn} studied the RG flow lines in the pure gauge SU(2) system. They found that near the extension of the first order phase transition line along the fundamental plaquette action  the RG flows away from the UVFP-2 and turns   around sharply at negative adjoint coupling. This  again  indicates that in this region the system is strongly influenced by the FP associated with the second order phase transition.
In a recent work \cite{Hasenfratz:inprep} we studied the pure gauge fundamental-adjoint SU(2) system with the 2-lattice matching MCRG method.
 Our results show that near the first order line and its extension towards negative adjoint couplings the MCRG matching method breaks down, the system is no longer in the basin of attraction of the perturbative G-FP.

When two UVFPs exist, numerical simulations have to stay in the vicinity of either one of them to describe the corresponding continuum physics.
Fortunately the basin of attraction of the UVFP-2  is at fairly strong coupling, and present day QCD lattice simulations are sufficiently far from it.    This, however, might not be the case in many fermion systems where interesting physics is expected to occur at strong gauge coupling. 

\begin{figure}
\vskip -.0cm
\begin{center}
\includegraphics[width=0.3\textwidth,clip]{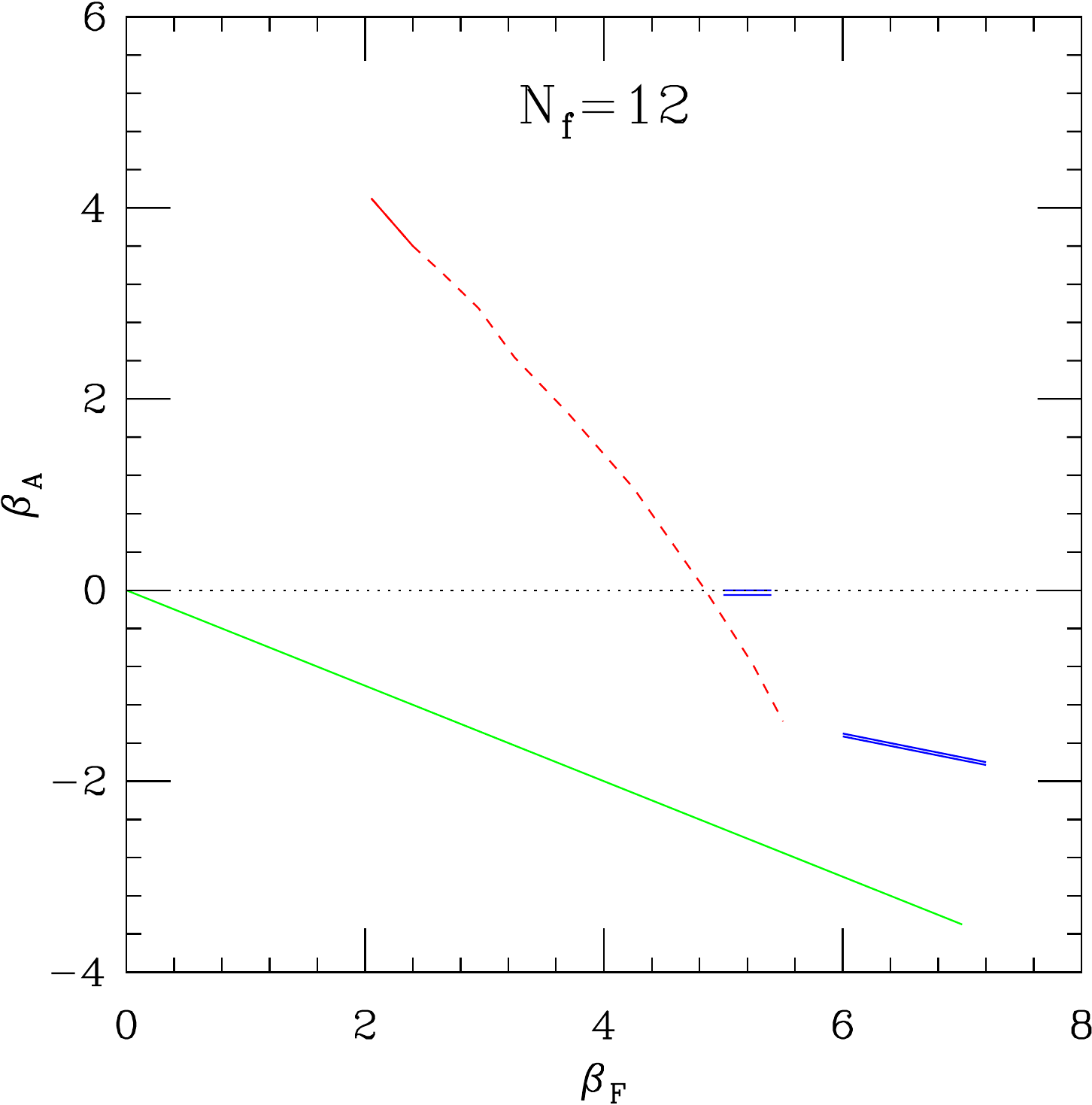}
\end{center}
\vskip -.15cm

\caption{ The approximate location of the phase transition /crossover in the fundamental-adjoint plane. The solid red line indicates first order phase transition while the dashed red line corresponds to crossovers.  
%The horizontal blue line at $\beta_A=0$ is the region of previous MCRG investigations \cite{Hasenfratz:2010fi}, while the blue line at $\beta_A/\beta_F=-0.25$ indicates the region studied in this work. The green line at $\beta_A/\beta_F=-0.5$  signals a change in the universality class. 
} 

\label{fig:phase_transition}

\end{figure}

Large number of fermions could change the phase structure of the pure gauge system so  I started by the study of the phase diagram of the fundamental-adjoint plaquette gauge action with 12 fermions. I used nHYP smeared staggered fermions \cite{Hasenfratz:2007rf} and measured the plaquette, the specific heat through the derivative of the plaquette, the Polyakov line and the chiral susceptibility on $8^3\times4$, $8^4$, $12^3\times4$ and $12^3\times6$ lattices. The specific heat gave a very clear signal for a first order transition, continuing along a crossover line, as indicated by the solid and dashed red lines in Figure \ref{fig:phase_transition}. This phase transition/crossover has no dependence on the temporal lattice size, it is a bulk feature of the system. In the crossover region the 
chiral susceptibility gives no signal at all and  I found no evidence for a finite temperature phase transition even at very strong gauge coupling.  Otherwise the 
 phase diagram of the 12 flavor system looks very similar to the pure gauge one. There is a first order line ending at a second order fixed point around $(\beta_F\approx2.4,\beta_A\approx3.6)$. At smaller $\beta_A$ there is a crossover that gets weaker with decreasing adjoint coupling. By $\beta_A=-1.4$ (the last point along the dashed red line) there is only a very weak signal left. I should note that I did not determine the phase transition with high precision - my goal was to establish the qualitative features of the phase diagram.

  %%%%%%%%%%%%%%%%%%%%%%%%%%%%%%%%%
%%%%%%%%%%%%%%%%%%%%%%%%%%%%%%%%%%

The horizontal blue line at $\beta_A=0$ Figure \ref{fig:phase_transition} shows the region I studied in Ref. \cite{Hasenfratz:2010fi}. MCRG matching became impossible at stronger couplings, to the left of the blue line. In retrospect that was most likely due to the nearby  crossover region. It is useful to recall that the leading order perturbative relation between the gauge coupling and the lattice couplings is
\bee
\frac{2N_c}{g^2}=\beta_F(1+2\frac{\beta_A}{\beta_F})\,.
\label{eq:beta_rel}
\ee
This suggests that the coupling $(\beta_F,\beta_A)$=(5.0,0.0) corresponds, at least perturbatively, to $(\beta_F,\beta_A)$=(10.0, -2.5). The latter point is quite far from the crossover along the $\beta_A/\beta_F=-0.25$ action line, indicated by the second blue line in Figure \ref{fig:phase_transition}. 
Finally, the green line in the figure corresponds to $\beta_A/\beta_F=-0.50$, the limit where the adjoint plaquette overtakes the fundamental one and flips the system into a new universality class. 

If the basin of attraction of the  perturbative G-FP is limited by the first order/crossover line, and Eq. \ref{eq:beta_rel} is any indication of constant physics, 
than along the $\beta_A/\beta_F= -0.25$ line one could reach considerably stronger couplings than with the $\beta_A=0$ fundamental action. 
I have chosen this action for the investigation described in this paper. This is a rather arbitrary choice, and other ratios could work equally well or even better.

The 2-lattice matching MCRG method is a powerful tool to numerically calculate the bare step scaling function, the discretized lattice analogue of the RG $\beta$ function. The method has been used for many years and recently it  has been discussed in detail in Refs.\cite{Hasenfratz:2009ea,Hasenfratz:2010fi}. 
Here I summarize it only briefly, concentrating on the new developments. 

The bare step scaling function is $s_b(\beta;s)=\beta -\beta'$, where $\beta$ and $\beta'$ are gauge couplings with lattice correlation lengths related as $\xi(\beta) = s \xi(\beta')$, where $s>1$ is an arbitrary scale parameter. In my work I always consider $s=2$ transformations and in the following I will drop the index $s$ in $s_b$. The lattice correlation length can be defined through different observables, which leads to systematical uncertainties in the definition of $s_b$. These can be controlled by approaching the UVFP, just like in the case of the renormalized step scaling function. In this work I will not perform this continuum extrapolation, rather include the systematical uncertainties  in the error estimate of the final result. The step scaling function approaches a constant at the G-FP, vanishes at  other fixed points (both UV and IR), and has the opposite sign of the RG $\beta$ function  where  it is non-zero.

  %%%%%%%%%%%%%%%%%%%%%%%%%%%%%%%%%
\begin{figure}
\vskip -.0cm
\begin{center}
\includegraphics[width=0.3\textwidth,clip]{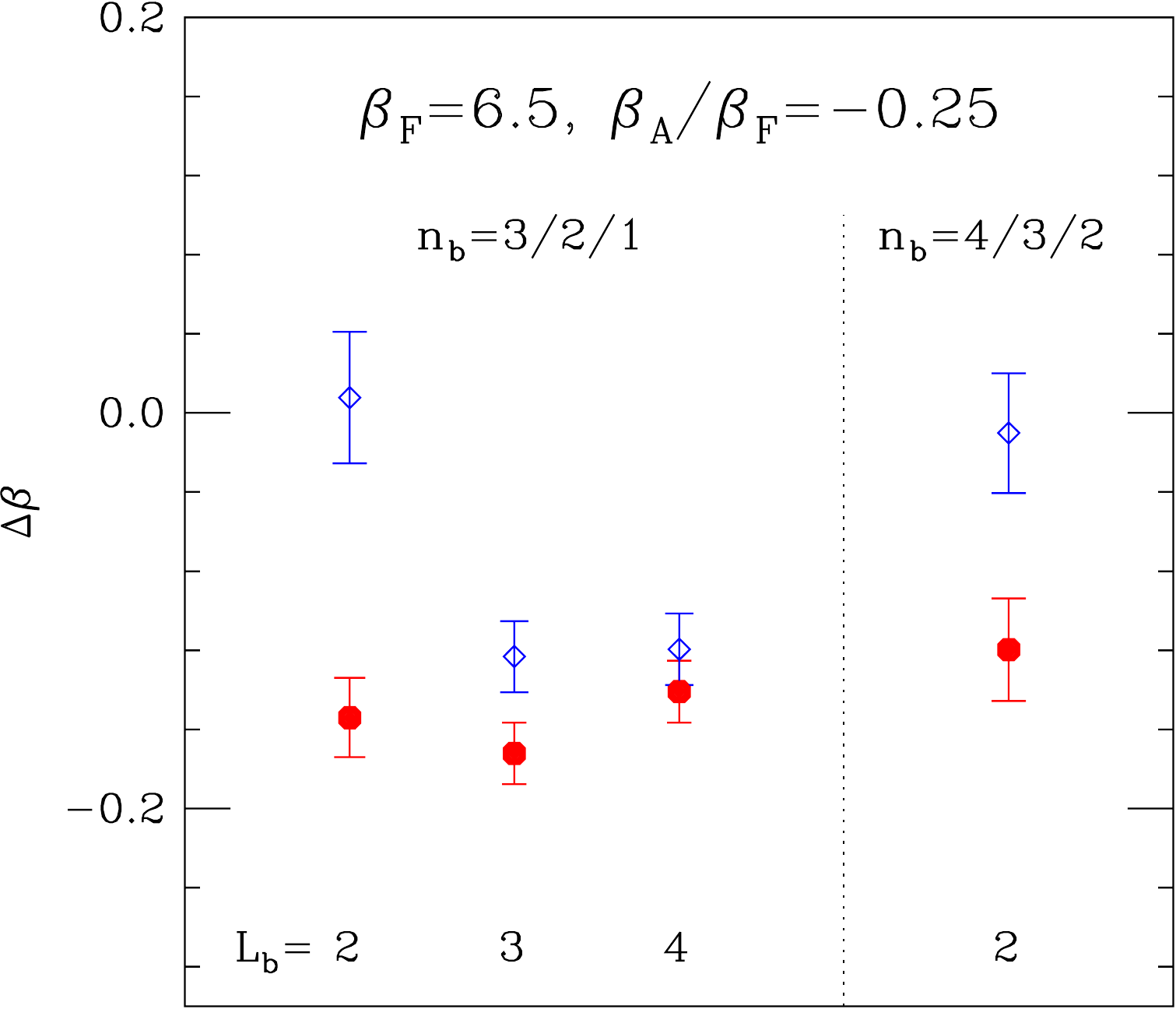}
\end{center}
\vskip -.15cm

\caption{ The optimized $\Delta\beta$ at $\beta_F=6.5$ with the action $\beta_A/\beta_F=-0.25$ . The red circles are the finite volume corrected predictions, the blue diamonds have no finite volume correction in the optimization step. The left side of the figure  shows results after comparing blocking steps $n_b=3/2/1$ on different volumes,  while right side is the result after one more blocking step. The data points are labeled by the final blocked lattice size $L_b$.  } 

\label{fig:finite_vol_b65}

\end{figure}
%%%%%%%%%%%%%%%%%%%%%%%%%%%%%%%%%%

The 2-lattice matching MCRG method relies on  matching  observables after several RG blocking steps. Its advantage is that the simulations do not have to be performed on volumes with lattice size  comparable or larger than the correlation length, and most of the finite size effects can be cancelled by comparing blocked observables measured on identical blocked volumes. 

Under repeated renormalization group blocking the action first flows towards the renormalized trajectory (RT), than along it, away from the UVFP. If the RG flow lines originating at $\beta$ and $\beta'$ hit the RT at the same points but require $n_b$ and $n_b-1$ blocking steps to do so, the requirement on the correlation length, $\xi(\beta) = 2 \xi(\beta')$, is satisfied. There are two steps to achieve this:
\begin{itemize}
\item {\it Matching:} If the blocked actions are identical, observables measured on $n_b$ times blocked configurations generated at $\beta$ have to match observables measured on $n_b-1$ times blocked configurations generated at $\beta'$. Since each blocking step 
reduces the lattice size by a factor of two, the
final  measurements are usually done on very small volumes. In order to minimize finite size effects it is best to do the simulations on twice as large lattices at $\beta$ than at $\beta'$ so the final measurements are done on the same blocked volume. Thus the shift in the gauge coupling is defined as $\Delta\beta_{\cal O}(\beta;n_b,L_b) = \beta -\beta'$ if 
\bee
\langle  \mathcal O(\beta;n_b,L_b) \rangle = \langle  \mathcal O(\beta';n_b-1,L_b)\rangle\,,
\label{eq:match}
\ee
where $\langle \mathcal O \rangle$ denotes the expectation value of some short distance operator and $L_b$ is the volume after $n_b$ and $n_b-1$ blocking steps (the same for both sides of Eq. \ref{eq:match}).
\item {\it Optimization:} The quantity $\Delta\beta$ defined in the previous step can differ significantly from the step scaling function $s_b$ if the RG flow does not reach the RT in $n_b-1$ steps. Most RG block transformations have a free parameter, usually denoted by $\alpha$, that can be optimized to minimize the number of RG steps needed to reach the RT. The optimized parameter is defined as the one where consecutive blocking steps predict the same shift, 
\bee
\Delta\beta_{\mathcal O}(\beta;n_b,L_b,\alpha_{\rm opt})= \Delta\beta_{\mathcal O}(\beta;n_b-1,L_b,\alpha_{\rm opt})\,.
\label{eq:optim}
\ee 
To minimize finite size effects $\Delta\beta_{\mathcal O}$ on the two sides of Eq. \ref{eq:optim} should be calculated on the same blocked lattice size $L_b$. Previous studies did not take this volume dependence   into account and usually satisfied Eq. \ref{eq:optim} on  different volumes. The error introduced this way is much smaller than the one  introduced by not matching the volume in Eq. \ref{eq:match}, but still it can be important when $\Delta\beta_{\mathcal O}$ itself is small.
\end{itemize}

Eq. \ref{eq:optim} requires the comparison of simulations on three different volumes. It is easiest to illustrate this with a specific example. Let's assume we simulate on $32^4$ volumes at some $\beta$ value. After blocking the lattices $n_b=3$ times we measure observables on $L_b=4$ lattices. We match these to observables measured on $n_b=2$ times  blocked lattices at some $\beta'$ coupling and find $\Delta\beta_{\mathcal O}(\beta;n_b=3,L_b=4)=\beta-\beta'$. Since $L_b=4$, simulations at $\beta'$ must have been done on $16^4$ lattices. Optimization requires that 
$$\Delta\beta_{\mathcal O}(\beta;n_b=3,L_b=4)=\Delta\beta_{\mathcal O}(\beta;n_b=2,L_b=4)\,.$$
 To calculate the quantity on the right hand side we have to do simulations on $16^4$ and $8^4$ volumes at $\beta$ and $\beta'$. 
 Identifying the optimal RG transformation and corresponding $\Delta\beta_{\mathcal O}$  with $n_b=3/2/1$ blocking steps requires simulations on volumes $32^4, 16^4$ and $8^4$. The procedure can be repeated with different $\mathcal O$ operators and the standard deviation between the predicted $\Delta\beta_{\mathcal O}$ values characterize the systematical errors of the matching. Results on larger volumes with more blocking levels provide further consistency checks.

The rest of this paper illustrates the optimization/finite size correction process and shows the step scaling function for a range of gauge couplings. The simulations were done on volumes between $32^4$ and $4^4$ using 12 nHYP smeared staggered fermions. The gauge action is a combination of fundamental and adjoint plaquette terms with fixed   $\beta_A/\beta_F=-0.25$ ratio. The lattice fermion masses were $am=0.0025$ on the $32^4$, 0.005 on the $16^4$ and $24^4$, 0.01 on the $12^4$ and $8^4$ and  0.02 on $6^4$ and  $4^4$ volumes.  The masses were chosen such that their values match  if they scale with their engineering dimension. This is not the right scaling if the anomalous mass happens to be large. However these bare fermion mass values are so small that the data show no mass dependence even with masses twice as large as used here. For all practical purposes these mass values can be considered to be in the chiral limit. I used an RG block transformation based on 2 HYP smearing steps with fixed inner parameters and considered 5 different operators as described in Ref.\cite{Hasenfratz:2010fi}

Figure \ref{fig:finite_vol_b65} illustrates the optimization at $\beta_F=6.5$. The left side of the figure shows the optimal $\Delta\beta$ after $n_b=3/2/1$ blocking steps and final blocked lattices of $L_b=$2, 3 and 4. As is easy to check, the original volumes in the matching sequences were $32\to16\to8$, $24\to12\to6$ and $16\to8\to4$. The red circles show the results of the optimized matching, a consistent value between all three volume series. The blue diamonds show the predicted $\Delta\beta$ without finite volume correction in the optimization. The result on the smallest volume set is clearly off, signaling  large finite volume effects. The two larger volumes show very little deviation, it appears that at least with my blocking transformations and 5 operators a finite volume of $L_b=3$ is already sufficient to minimize these second order finite volume effects. The right side of the figure shows $\Delta\beta$ after one more blocking step, with $n_b=4/3/2$. The largest volume in this case was $32^4$ with the final blocked volume $L_b=2$. Again, the finite volume corrected optimized data is significantly different from the uncorrected one but both are consistent with the $L_b=2$ results of the left hand side. The error bars on the data points come from a combination of statistical and systematical errors. They are dominated by systematical errors in the $n_b=3/2/1$ sequence and by statistical errors in the $n_b=4/3/2$ one.  Comparing the finite volume corrected optimized results for  $\Delta\beta$  on all three volume sequences and after $n_b=3/2/1$ and $n_b=4/3/2$ blocking levels one finds $s_b(\beta=6.5)=-0.15(2)$ .

 %%%%%%%%%%%%%%%%%%%%%%%%%%%%%%%%%
\begin{figure}
\vskip -.0cm
\begin{center}
\includegraphics[width=0.3\textwidth,clip]{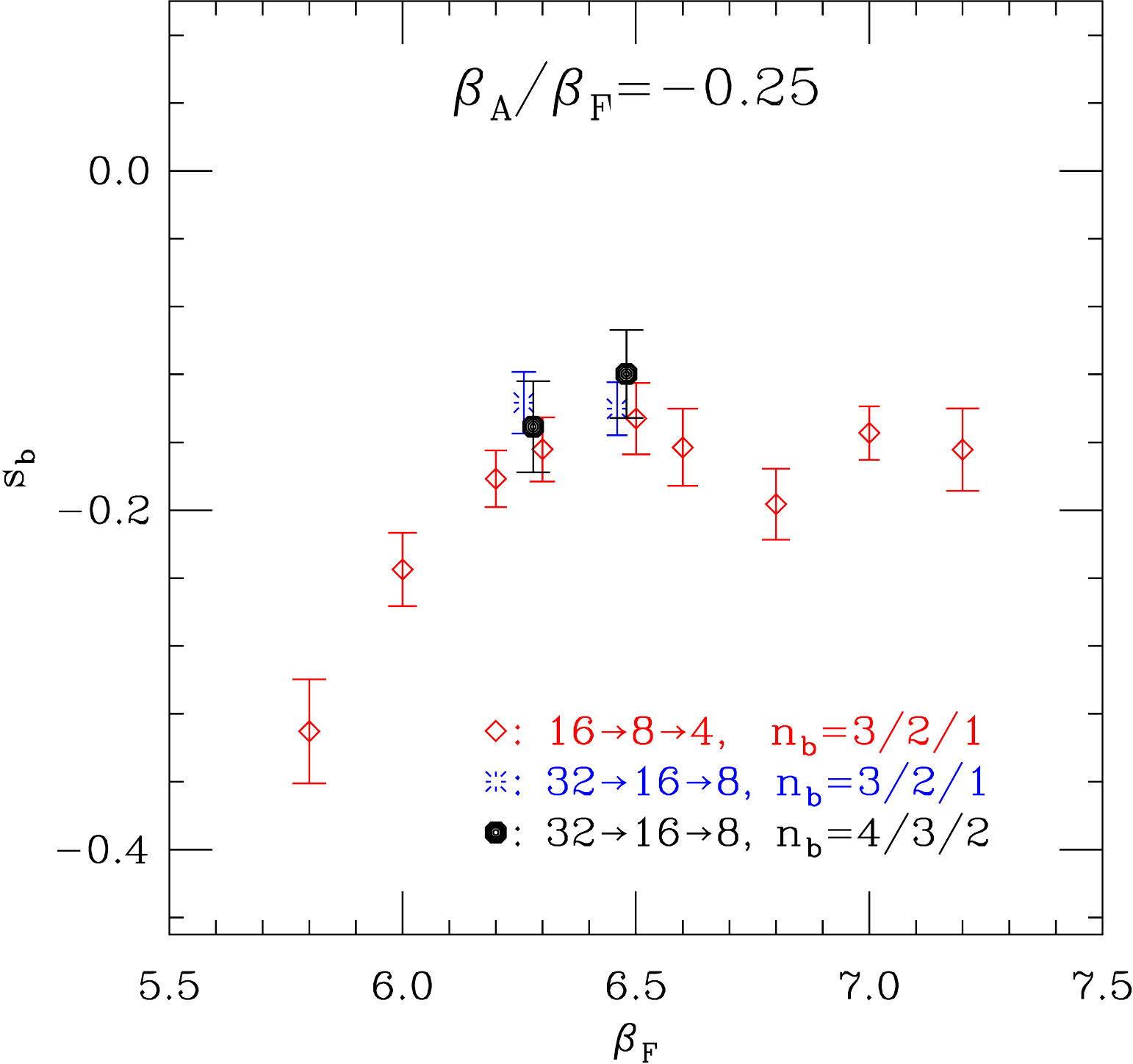}
\end{center}
\vskip -.15cm

\caption{ The bare step scaling function.
%or 12 flavors of fundamental fermions with a gauge action containing fundamental and adjoint plaquettes with fixed ratio $\beta_A/\beta_F=-0.25$. 
The different  symbols correspond to predictions from optimized matching on different lattice volumes and blocking levels.  } 

\label{fig:db_all}

\end{figure}
%%%%%%%%%%%%%%%%%%%%%%%%%%%%%%%%%%

 Figure \ref{fig:db_all} show $\Delta\beta$ (or $s_b$) at a range of gauge couplings. The red diamond points are from $16\to8\to4$,  $n_b=3/2/1$, the blue crosses are from  $32\to16\to8$,  $n_b=3/2/1$,  and the black circles are from $32\to16\to8$, $n_b=4/3/2$ optimized matching. Where all three data points are available, they are consistent. Overall, the data show that $s_b$ is negative in the investigated region, indicating that the RG $\beta$ function has crossed zero and the measurements are on the strong coupling side of the IRFP. One should note that the data in Figure \ref{fig:db_all} does not correspond to any given RG transformation. Each point has a slightly different optimization parameter and can have a different IRFP as well.

In summary, Figure \ref{fig:db_all} gives strong evidence that the 12 fermion SU(3) system is infrared conformal. This is not the first MCRG investigation of this theory, but previous ones were inconclusive. The success this time had two sources. I considered an action farther away from a secondary UVFP caused by strong coupling lattice artifacts and that made simulations possible at physically stronger gauge couplings. Second I corrected for a previously ignored finite volume effect that made the results obtained on different volumes after different blocking levels consistent. This finite volume correction also reduced the systematical errors that come from matching 5 different operators. The same approach could easily be applied to other models near the conformal window.

The numerical calculations of this work were carried out on the HEP-TH computer cluster at the University of Colorado and at FNAL, under the auspices of USQCD and SciDAC. 
This research was partially supported by the US Department of Energy.

\bibliographystyle{apsrev}
\bibliography{lattice}

\end{document}